# Accurate and reduced SISO Sequence Impedance Models of Grid-tied Voltage Source Converter for Small Signal Stability Analysis

Chen Zhang[1], Xu Cai[1], Atle Rygg[2] and Marta Molinas[2]
1 Department of Wind Power Research Center, Shanghai Jiao Tong University, Shanghai, China.
2 Department of Engineering Cybernetics, Norwegian University of Science and Technology, Trondheim, Norway.

***Abstract*—Impedance models are widely used in assessing small signal stability of grid-tied voltage source converter (VSC) systems. Recent research has proven that impedance models of grid-tied VSC in both *dq* and sequence domains are generally Multi-Input Multi-Output (MIMO) systems, and the generalized Nyquist criterion has to be applied for stability analysis to these MIMO systems. However, finding Single-Input and Single-Output (SISO) equivalents for this system is always appealing because of the simplicity and the convenience for physical interpretation when assessing the stability, compared to MIMO systems. This paper presents two types of SISO impedance models of grid-tied VSC system, one is derived from the strong grid assumption, and the other is from the closed-loop equivalence. The accuracy of these models is assessed with respect to the measured impedances in PSCAD/EMTDC, and their effects on the stability assessment were analyzed as well. It is proven that the accurate SISO model gives identical result as the MIMO (matrix-based) impedance model with respect to stability analysis. However, the reduced SISO model may lead to wrong results if the bandwidth of phase locked loop is large.**

## I. INTRODUCTION

Nowadays, voltage source converters (VSC) have been widely used in the grid-integration of renewable energies [1] as well as the flexible power transmission system [2]. Oscillations at both low frequencies [3] and high frequencies [4] have been observed in VSC-based systems, in particular in weak grid conditions [5]. Such kinds of small signal stability issues can be effectively assessed by impedance-based analysis approach. Impedance models of e.g., three-phase VSC [6]-[9], single-phase VSCs [10], and multi-modular-converters [11] etc., have been developed rigorously in recent literatures. Although impedances are generally a description of the input-output characteristics in frequency domain, the exact formulation can be different due to the modeling techniques used.

For a typical two-level and three-phase grid-tied VSC, the impedances can be extracted either in *dq* synchronously rotating frame [8] or in three phase stationary frame [7]. In *dq* frame, the grid-tied VSC system is time-invariant if the three-phase grid is balanced. It allows linearization directly, thereby performing Laplace transformation on the resultant linear time invariant (LTI) model yields the *dq* impedances [9]. However, if seen from three phase stationary frame, the grid-tied VSC is time-varying inherently. Therefore, the *harmonic linearization method* in [12] has to be applied to obtain the sequence impedances [7]. Or more generally, by linearizing the time-varying system along a steady periodic trajectory yields a linear time periodic (LTP) system. In order to represent the LTP systems in a frequency domain framework as the LTI systems, the generalized *harmonic balance approach* [13] should be adopted.

Despite the different models in *dq* and sequence domain, both of them are coupled due to the presence of nonzero off diagonal terms in the impedance matrices. Therefore, recent researches have shown a great interest in the interpretation of these couplings and their consequences when assessing the stability. In [14] and [15], frequency couplings were identified in sequence domain, i.e. positive and negative sequence are coupled and separated by twice fundamental frequency and their impacts on low frequency stability were emphasized. This interesting property of VSC was also identified from *dq* impedance and introduced as *dq* asymmetry in [16]. Due to this implicit binding between *dq* and sequence impedances, the relationship between sequence impedance and *dq* impedance was further investigated in [17], and they identified that *dq* impedances can be transformed into modified sequence impedances by means of symmetrical decomposition method [18]. Similarly, the *complex space vector method* [16] was used to derive the VSC impedance in stationary $\alpha\beta$ frame directly from *dq* frame in [19] lately.

However, the frequency couplings in the foregoing review are in single frequency coupling form, i.e. a single frequency perturbation induces a single frequency coupling separated by twice fundamental frequency. This condition is true if either the converter or grid impedance is *dq asymmetric* [16] or the *mirror frequency coupling* effect in [17]. But if the asymmetry was presented in phase domain, e.g., asymmetric three phase grid, there tends to be multiple frequency couplings between VSCs and grid. In order to include these couplings *harmonic state space method* [20] can be adopted, more often the truncated harmonic transfer function for numeric and stability study in [13] was used.

At present, both the single and multiple frequency couplings cases can only be captured by matrix-based impedances which are multi-input and multi-output (MIMO) system by nature, therefore the generalized Nyquist criterion (GNC) [21] should be adopted for stability analysis, which obscures the advantages of applying sequence impedances. On the other hand, finding the single-input and single-output (SISO) equivalents of grid-tied VSC are always appealing, due to the simplicity and convenience for physical interpretation. This paper aims to achieve this objective by exploring further the properties of single frequency couplings from a physical point of view. The rest of the paper is organized as follows: Section II, the sequence impedance model of grid-tied VSC was developed from the knowledge of *dq* impedances. In section III, the system block diagrams in form of complex transfer function was established. According to the block diagrams, the properties of sequence couplings can be revealed intuitively. Subsequently two SISO sequence impedance models of grid-tied VSC with different accuracies were established. These models were compared with the measured impedances in PSCAD/EMTDC. Section IV performs the stability analysis using the proposed SISO models, the accuracy in predicting stability were analyzed both by numerical and time domain simulations. Section V draws the final conclusion.

## II. Modeling of grid-tied VSC in (modified) sequence domain

### A. Topology and control scheme of grid-tied VSC

Fig.1 presents the grid-tied VSC system to be analyzed in this paper. It constitutes a typical two-level VSC, a *L-type* filter and a Thevenin equivalent grid. For the control systems, only current controller and the phase-locked-loop are considered for the simplicity in the forthcoming property analysis. Note that, the voltage feed-forward terms are not shown in the current control loop. If the bandwidths of these feed-forward terms are not carefully chosen, they can affect both the transient [6] and small signal response [5] of VSC. In this regard, the feed-forward terms are viewed as impedance shaping effects and will not be discussed in this paper since our focus is on modeling.

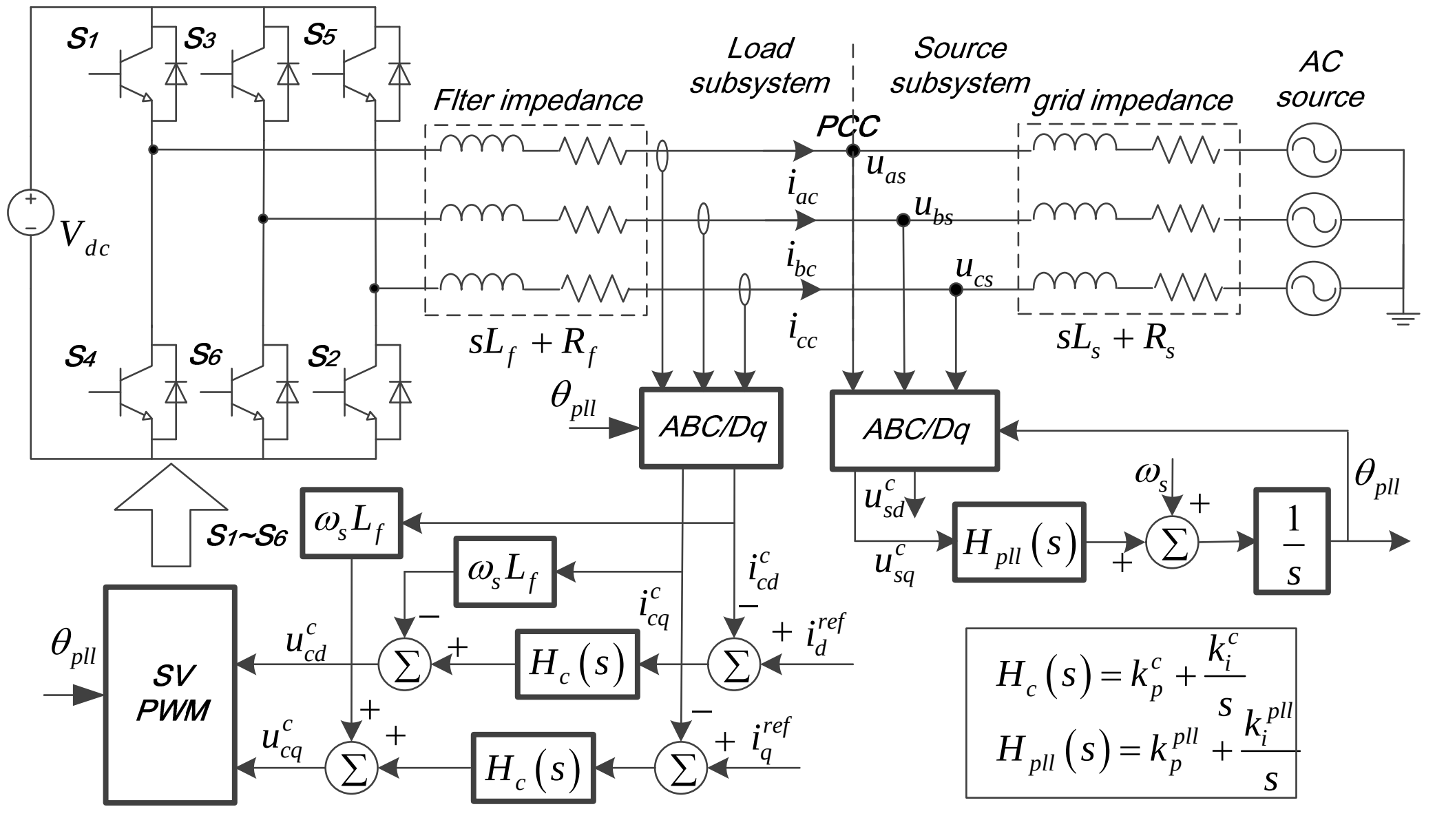

Fig. 1 Schematic of the grid-tied VSC system

B. Symmetrical decomposition of *dq* impedance

Taking a *dq* impedance model in [9] as an example:

$$\begin{bmatrix} U_d(s) \\ U_q(s) \end{bmatrix} = \begin{bmatrix} Z_{dd}(s) & Z_{dq}(s) \\ Z_{qd}(s) & Z_{qq}(s) \end{bmatrix} \begin{bmatrix} I_d(s) \\ I_q(s) \end{bmatrix} \quad (1),$$

Since (1) is a LTI system, a complex exponential input (e.g. $e^{st}$ ) will give an output in the same form [13], thus variables such as *dq* currents and voltages in (1) can be written explicitly with variable "*s*" as below.

$$\begin{bmatrix} \mathbf{U}_d \\ \mathbf{U}_q \end{bmatrix} e^{st} = \begin{bmatrix} Z_{dd}(s) & Z_{dq}(s) \\ Z_{qd}(s) & Z_{qq}(s) \end{bmatrix} \begin{bmatrix} \mathbf{I}_d \\ \mathbf{I}_q \end{bmatrix} e^{st}, \forall s \to j\omega \quad (2)$$

Where $s \to j\omega$ is a translation from *s*-domain to frequency-domain. $\mathbf{I}_d, \mathbf{I}_q$ and $\mathbf{U}_d, \mathbf{U}_q$ are the Fourier coefficients of currents and voltages at frequency $\omega$. They can be further decomposed into positive and negative components as:

$$\begin{bmatrix} \mathbf{U}_p \\ \mathbf{U}_n \end{bmatrix} = \frac{1}{2} \begin{bmatrix} 1 & j \\ 1 & -j \end{bmatrix} \begin{bmatrix} \mathbf{U}_d \\ \mathbf{U}_q \end{bmatrix} = \mathbf{A} \begin{bmatrix} \mathbf{U}_d \\ \mathbf{U}_q \end{bmatrix} \quad (3)$$

Applying the matrix $\mathbf{A}$ and its inverse $\mathbf{A}^{-1}$ to (2) yields:

$$\begin{bmatrix} \mathbf{U}_p \\ \mathbf{U}_n \end{bmatrix} = \mathbf{A} \begin{bmatrix} Z_{dd}(s) & Z_{dq}(s) \\ Z_{qd}(s) & Z_{qq}(s) \end{bmatrix} \mathbf{A}^{-1} \begin{bmatrix} \mathbf{I}_p \\ \mathbf{I}_n \end{bmatrix} = \mathbf{Z}_{PN}(s) \begin{bmatrix} \mathbf{I}_p \\ \mathbf{I}_n \end{bmatrix}, \forall s \to j\omega \quad (4)$$

Where each element in $\mathbf{Z}_{PN}(s) = \begin{bmatrix} \mathbf{Z}_{pp}(s) & \mathbf{Z}_{pn}(s) \\ \mathbf{Z}_{np}(s) & \mathbf{Z}_{nn}(s) \end{bmatrix}$ is generally a complex transfer function. This formula has been rigorously developed in [17] by the same authors as this paper. Additionally, a similar impedance model was derived in [19] using complex space vector method.

C. Frequency notation specifications

The frequency notation of impedances can be different due to different modeling techniques used. This section will clarify the frequency notation used in [14], [17] and [19]. Essentially, the frequency notation in this paper is the same as [17], i.e. the frequency (e.g. in (4)) is referred to the *dq* frame and is always positive. This is due to fact that symmetrical decomposition method only utilizes the phasors of positive frequency. Therefore, the term *modified* refers to the frequency notation in this work.

In [14], a different frequency notation is used, which is referred to the stationary frame or phase domain. However, the variable "s" in [14] (denoted by $s_{[14]}$) can be related with the variable "*s*" in this paper (denoted by $s_{\mathrm{mod}}$) by frequency shift, e.g. $s_{\mathrm{mod}} = s_{[14]} - j\omega_1$ for the positive sequence and $s_{\mathrm{mod}} = s_{[14]} + j\omega_1$ for negative sequence. Substitute $s_{\mathrm{mod}}$ into (4) (admittance is used for comparison) yields:

$$\begin{aligned}
\begin{bmatrix} \mathbf{I}_p\left(s_{[14]}\right) \\ \mathbf{I}_n\left(s_{[14]} - j2\omega_1\right) \end{bmatrix} &= \begin{bmatrix} \mathbf{Y}_{pp}\left(s_{[14]} - j\omega_1\right) \\ \mathbf{Y}_{np}\left(s_{[14]} - j\omega_1\right) \end{bmatrix} \mathbf{U}_p\left(s_{[14]}\right) = \begin{bmatrix} Y_p\left(s_{[14]}\right) \\ J_p\left(s_{[14]}\right) \end{bmatrix} \mathbf{U}_p\left(s_{[14]}\right), Pos \\
\begin{bmatrix} \mathbf{I}_p\left(s_{[14]} + j2\omega\right) \\ \mathbf{I}_n\left(s_{[14]}\right) \end{bmatrix} &= \begin{bmatrix} \mathbf{Y}_{pn}\left(s_{[14]} + j\omega_1\right) \\ \mathbf{Y}_{nn}\left(s_{[14]} + j\omega_1\right) \end{bmatrix} \mathbf{U}_n\left(s_{[14]}\right) = \begin{bmatrix} J_n\left(s_{[14]}\right) \\ Y_n\left(s_{[14]}\right) \end{bmatrix} \mathbf{U}_n\left(s_{[14]}\right), Neg
\end{aligned} \tag{5}$$

Shifting the frequency of negative sequence component by $s_{[14]} = s_{[14]} - j2\omega$ yields:

$$\begin{bmatrix} \mathbf{I}_p\left(s_{[14]}\right) \\ \mathbf{I}_n\left(s_{[14]} - j2\omega_1\right) \end{bmatrix} = \begin{bmatrix} Y_p\left(s_{[14]}\right) & J_p\left(s_{[14]} - j2\omega_1\right) \\ J_p\left(s_{[14]}\right) & Y_n\left(s_{[14]} - j2\omega_1\right) \end{bmatrix} \begin{bmatrix} \mathbf{U}_p\left(s_{[14]}\right) \\ \mathbf{U}_n\left(s_{[14]} - j2\omega_1\right) \end{bmatrix} \tag{6}$$

As a result, the admittance in (6) has the same structure as in 错误!未找到引用源。, which shows that the (modified) sequence impedance can be related to the phase domain sequence impedance via frequency notation translation.

On the other hand, a complex-space-vector approach in [19] was adopted, which leads to another frequency notation. It allows the variable "s" to be negative for the representation of negative sequences or reversely rotated space vectors. However, the model in [19] can also be obtained using the model in this paper, e.g., suppose a complex space vector in stationary frame as $\mathbf{I}(s) = \mathbf{I} \cdot e^{st}$, the resulting sequence components using (4) is $\mathbf{U}_p = \mathbf{Z}_{pp}\left(s - j\omega_1\right)\mathbf{I}$ and $\mathbf{U}_n = \mathbf{Z}_{np}\left(s - j\omega_1\right)\mathbf{I}$, hence the vector in *dq* frame can be derived as $\mathbf{v}_{dq} = \mathbf{U}_p e^{(s - j\omega_1)t} + \mathbf{U}_n^* e^{(\bar{s} + j\omega_1)t}$, while in stationary frame, the vector is $\mathbf{v} = \mathbf{v}_{dq} e^{j\omega_1 t} = \mathbf{U}_p e^{st} + \mathbf{U}_n^* e^{(\bar{s} + j2\omega_1)t}$. Substitute the sequence phasor into the vectors yields:

$$\mathbf{v} = \mathbf{Z}_{pp}\left(s - j\omega_1\right)\mathbf{I}\cdot e^{st} + \mathbf{Z}_{np}^{*}\left(\overline{s} + j\omega_1\right)\mathbf{I}^{*}\cdot e^{\left(\overline{s}+j2\omega_1\right)t} \quad (7)$$

Conjugating the vector in (7) yields:

$$e^{j2\omega_1 t}\mathbf{v}^{*} = \mathbf{Z}_{pp}^{*}\left(\overline{s} + j\omega_1\right)\mathbf{I}^{*}\cdot e^{\left(\overline{s}+j2\omega_1\right)t} + \mathbf{Z}_{np}\left(s - j\omega_1\right)\mathbf{I}\cdot e^{st} \quad (8)$$

Rewriting (7) and (8) in a compact matrix form yields:

$$\begin{bmatrix} \mathbf{v} \\ e^{j2\omega_1 t}\mathbf{v}^{*} \end{bmatrix} = \begin{bmatrix} \mathbf{Z}_{pp}\left(s - j\omega_1\right) & \mathbf{Z}_{np}^{*}\left(\overline{s} + j\omega_1\right) \\ \mathbf{Z}_{np}\left(s - j\omega_1\right) & \mathbf{Z}_{pp}^{*}\left(\overline{s} + j\omega_1\right) \end{bmatrix} \begin{bmatrix} \mathbf{I} \\ e^{j2\omega_1 t}\mathbf{I}^{*} \end{bmatrix} \quad (9)$$

Note that the complex conjugate operator "*" denotes conjugation of the function, whereas the upper line "—" denotes conjugation of the variable. For the complex transfer functions in (4), $\mathbf{Z}_{np}^{*}\left(s\right) = \mathbf{Z}_{pn}\left(s\right)$, $\mathbf{Z}_{pp}^{*}\left(s\right) = \mathbf{Z}_{nn}\left(s\right)$. Therefore, (9) has the same structure as in [19], which in this paper is derived from the (modified) sequence impedance with carefully examined frequency notation.

In general, the above models are related via linear transformations. Therefore, stability analysis should be identical in all of these models.

D. System blocks of grid-tied VSC in sequence domain

Applying the transformation method in Section II.B, the system blocks of grid tied-VSC in (modified) sequence domain is shown bellow:

Table 1 *dq* [3] and sequence impedance blocks

| Blocks | Variables | | *dq* domain | Variables | | Sequence domain | Circuit Eqv. |
|---|---|---|---|---|---|---|---|
| | input | output | | Input | output | | |
| Control | $\begin{bmatrix} i_{cd}^{s} \\ i_{cq}^{s} \end{bmatrix}$ | $\begin{bmatrix} u_{cd}^{s} \\ u_{cq}^{s} \end{bmatrix}$ | $\begin{bmatrix} H_c(s) & \\ & H_c(s) \end{bmatrix}$ | $\begin{bmatrix} \mathbf{i}_c^{p} \\ \mathbf{i}_c^{n} \end{bmatrix}$ | $\begin{bmatrix} \mathbf{u}_c^{p} \\ \mathbf{u}_c^{n} \end{bmatrix}$ | $\begin{bmatrix} \mathbf{H}_c^{pp}(s) & \\ & \mathbf{H}_c^{nn}(s) \end{bmatrix}$ | 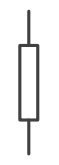 |
| Feed-forward | $\begin{bmatrix} i_{cd}^{s} \\ i_{cq}^{s} \end{bmatrix}$ | $\begin{bmatrix} u_{cd}^{s} \\ u_{cq}^{s} \end{bmatrix}$ | $\begin{bmatrix} & -\omega_s L_f \\ \omega_s L_f & \end{bmatrix}$ | $\begin{bmatrix} \mathbf{i}_c^{p} \\ \mathbf{i}_c^{n} \end{bmatrix}$ | $\begin{bmatrix} \mathbf{u}_c^{p} \\ \mathbf{u}_c^{n} \end{bmatrix}$ | $\begin{bmatrix} j\omega_s L_f & \\ & -j\omega_s L_f \end{bmatrix}$ | 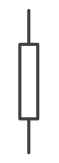 |
| PLL | $\begin{bmatrix} u_{sd}^{s} \\ u_{sq}^{s} \end{bmatrix}$ | $\begin{bmatrix} i_{cd}^{c} \\ i_{cq}^{c} \end{bmatrix}$ | $\begin{bmatrix} 0 & \frac{I_{cq0}T_{pll}(s)}{V_0} \\ 0 & -\frac{I_{cd0}T_{pll}(s)}{V_0} \end{bmatrix}$ | $\begin{bmatrix} \mathbf{u}_s^{p} \\ \mathbf{u}_s^{n} \end{bmatrix}$ | $\begin{bmatrix} \mathbf{i}_c^{p} \\ \mathbf{i}_c^{n} \end{bmatrix}$ | $\frac{T_{pll}(s)}{2V_0}\begin{bmatrix} -\mathbf{I}_{c0} & \mathbf{I}_{c0} \\ \mathbf{I}_{c0}^{*} & -\mathbf{I}_{c0}^{*} \end{bmatrix}$ | 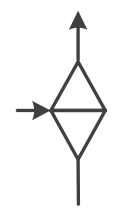 |
| | $\begin{bmatrix} u_{sd}^{s} \\ u_{sq}^{s} \end{bmatrix}$ | $\begin{bmatrix} u_{cd}^{s} \\ u_{cq}^{s} \end{bmatrix}$ | $\begin{bmatrix} 0 & -\frac{V_{cq0}T_{pll}(s)}{V_0} \\ 0 & \frac{V_{cd0}T_{pll}(s)}{V_0} \end{bmatrix}$ | $\begin{bmatrix} \mathbf{u}_s^{p} \\ \mathbf{u}_s^{n} \end{bmatrix}$ | $\begin{bmatrix} \mathbf{u}_c^{p} \\ \mathbf{u}_c^{n} \end{bmatrix}$ | $\frac{T_{pll}(s)}{2V_0}\begin{bmatrix} \mathbf{V}_{c0} & -\mathbf{V}_{c0} \\ -\mathbf{V}_{c0}^{*} & \mathbf{V}_{c0}^{*} \end{bmatrix}$ | 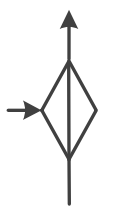 |
| Filter | $\begin{bmatrix} i_{cd}^{s} \\ i_{cq}^{s} \end{bmatrix}$ | $\begin{bmatrix} u_{cd}^{s} \\ u_{cq}^{s} \end{bmatrix}$ | $\begin{bmatrix} Z_f(s) & -\omega_s L_f \\ \omega_s L_f & Z_f(s) \end{bmatrix}$ | $\begin{bmatrix} \mathbf{i}_c^{p} \\ \mathbf{i}_c^{p} \end{bmatrix}$ | $\begin{bmatrix} \mathbf{u}_c^{p} \\ \mathbf{u}_c^{n} \end{bmatrix}$ | $\begin{bmatrix} \mathbf{Z}_f^{pp}(s) & \\ & \mathbf{Z}_f^{nn}(s) \end{bmatrix}$ | 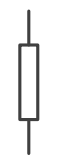 |
| Grid | $\begin{bmatrix} i_{sd}^{s} \\ i_{sq}^{s} \end{bmatrix}$ | $\begin{bmatrix} u_{sd}^{s} \\ u_{sq}^{s} \end{bmatrix}$ | $\begin{bmatrix} Z_s(s) & -\omega_s L_s \\ \omega_s L_s & Z_s(s) \end{bmatrix}$ | $\begin{bmatrix} \mathbf{i}_s^p \\ \mathbf{i}_s^p \end{bmatrix}$ | $\begin{bmatrix} \mathbf{u}_s^p \\ \mathbf{u}_s^n \end{bmatrix}$ | $\begin{bmatrix} \mathbf{Z}_S^{pp}(s) & \\ & \mathbf{Z}_S^{nn}(s) \end{bmatrix}$ | 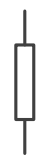 |

In table.1, $\mathbf{I}_{c0} = I_{cd0} + jI_{cq0}$ and $\mathbf{V}_{c0} = V_{cd0} + jV_{cq0}$ are current and voltage phasors of VSC at steady state respectively. $V_0$ is the nominal voltage at PCC. $Z_f(s) = sL_f + R_f$ , $Z_s(s) = sL_s + R_s$ , $T_{pll}(s) = \dfrac{V_0 H_{pll}(s)}{s + V_0 H_{pll}(s)}$, $\mathbf{Z}_f^{pp}(s) = Z_f(s) + j\omega_s L_f$ , $\mathbf{Z}_f^{nn}(s) = Z_f(s) - j\omega_s L_f$ , $\mathbf{Z}_S^{pp}(s) = Z_s(s) + j\omega_s L_s$ , $\mathbf{Z}_S^{nn}(s) = Z_s(s) - j\omega_s L_s$ . $\mathbf{H}_c^{pp}(s) = \mathbf{H}_c^{nn}(s) = H_c(s)$ .The superscript "s" for *dq* variables denote system domain, whereas the superscript "c" denotes converter domain [9]. Subscripts "s" and "c" denote variables in the grid side and converter side respectively. Transfer function in bold type denotes complex transfer functions.

E. Sequence impedance model of grid-tied VSC system

Combining the sequence transfer function blocks in Table.1 in accordance with system schematics in Fig.1, the VSC (Load) admittance and grid (Source) impedance can be written as (generation notation is used for the current direction):

$$-\begin{bmatrix} \mathbf{i}_L^p \\ \mathbf{i}_L^n \end{bmatrix} = \begin{bmatrix} \mathbf{Y}_L^{pp}(s) & \mathbf{Y}_L^{pn}(s) \\ \mathbf{Y}_L^{np}(s) & \mathbf{Y}_L^{nn}(s) \end{bmatrix} \begin{bmatrix} \mathbf{u}_L^p \\ \mathbf{u}_L^n \end{bmatrix} = \mathbf{Y}_L^{PN}(s) \begin{bmatrix} \mathbf{u}_L^p \\ \mathbf{u}_L^n \end{bmatrix} \quad (10)$$

$$\begin{bmatrix} \mathbf{u}_S^p \\ \mathbf{u}_S^n \end{bmatrix} = \begin{bmatrix} \mathbf{Z}_S^{pp} & 0 \\ 0 & \mathbf{Z}_S^{nn} \end{bmatrix} \begin{bmatrix} \mathbf{i}_S^p \\ \mathbf{i}_S^n \end{bmatrix} = \mathbf{Z}_S^{PN} \begin{bmatrix} \mathbf{i}_S^p \\ \mathbf{i}_S^n \end{bmatrix} \quad (11)$$

$$\mathbf{i}_S^p = \mathbf{i}_L^p, \mathbf{i}_S^n = \mathbf{i}_L^n \quad (12)$$

$$\mathbf{u}_S^p + \mathbf{u}_{inj}^p = \mathbf{u}_L^p, \mathbf{u}_L^n = \mathbf{u}_S^n \quad (13)$$

Where $\mathbf{G}_{pll}(s) = \dfrac{T_{pll}(s)}{2V_0}\left(H_c(s)\mathbf{I}_{c0} + \mathbf{V}_{c0}\right)$, $\mathbf{Y}_L^{pp}(s) = \dfrac{1 - \mathbf{G}_{pll}(s)}{\mathbf{H}_c^{pp}(s) + \mathbf{Z}_f^{pp}(s)}$, $\mathbf{Y}_L^{nn}(j\omega) = \left(\mathbf{Y}_L^{pp}(-j\omega)\right)^*$ , $\mathbf{Y}_L^{pn}(s) = \dfrac{\mathbf{G}_{pll}(s)}{\mathbf{H}_c^{pp}(s) + \mathbf{Z}_f^{pp}(s)}$, $\mathbf{Y}_L^{np}(j\omega) = \left(\mathbf{Y}_L^{pn}(j\omega)\right)^*$ . Note that, the superscript "PN" in capital form denotes matrix, whereas the "*pn*" in lower-case denotes elements in the matrix.

The sequence equivalent circuits can be draw in accordance with (4)-(8):

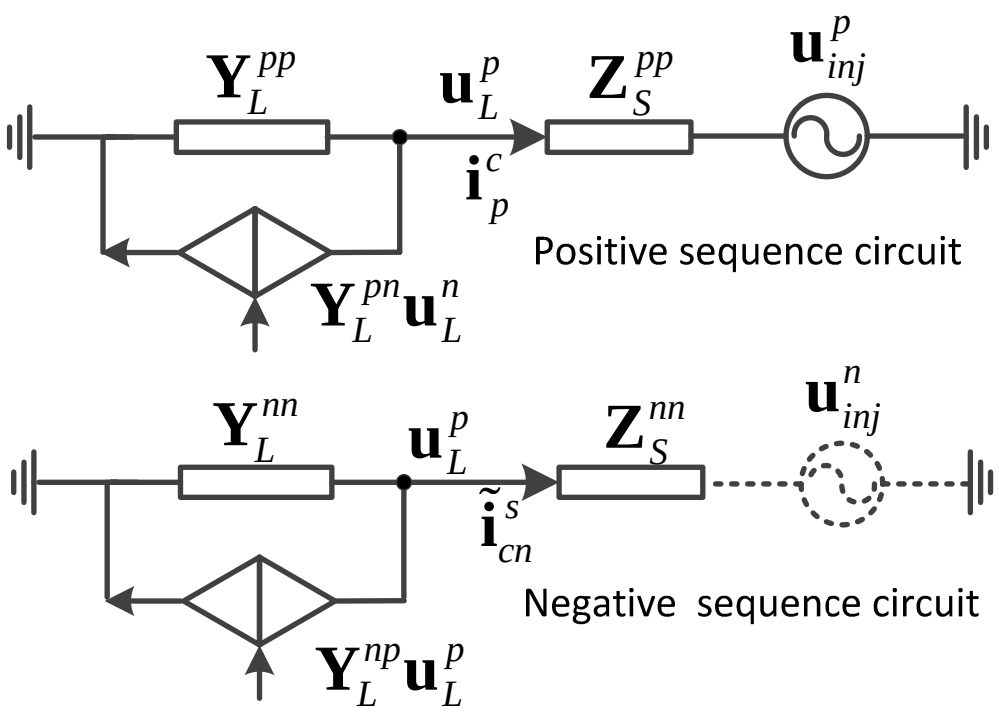

Fig. 2 Sequence equivalent circuits of grid-tied VSC system

At this point, the derived sequence impedance model can be used to assess stability with the help of GNC [14]. It has been proven in [17] that this analysis gives identical results as the GNC based on *dq* domain impedance.

III. SISO equivalents of grid-tied VSC system

A. Analysis of coupled sequence loops

Combining the sequence domain control blocks in Table 1 in accordance with the system configuration in Fig.1, yields sequence domain control blocks as shown in Fig.3:

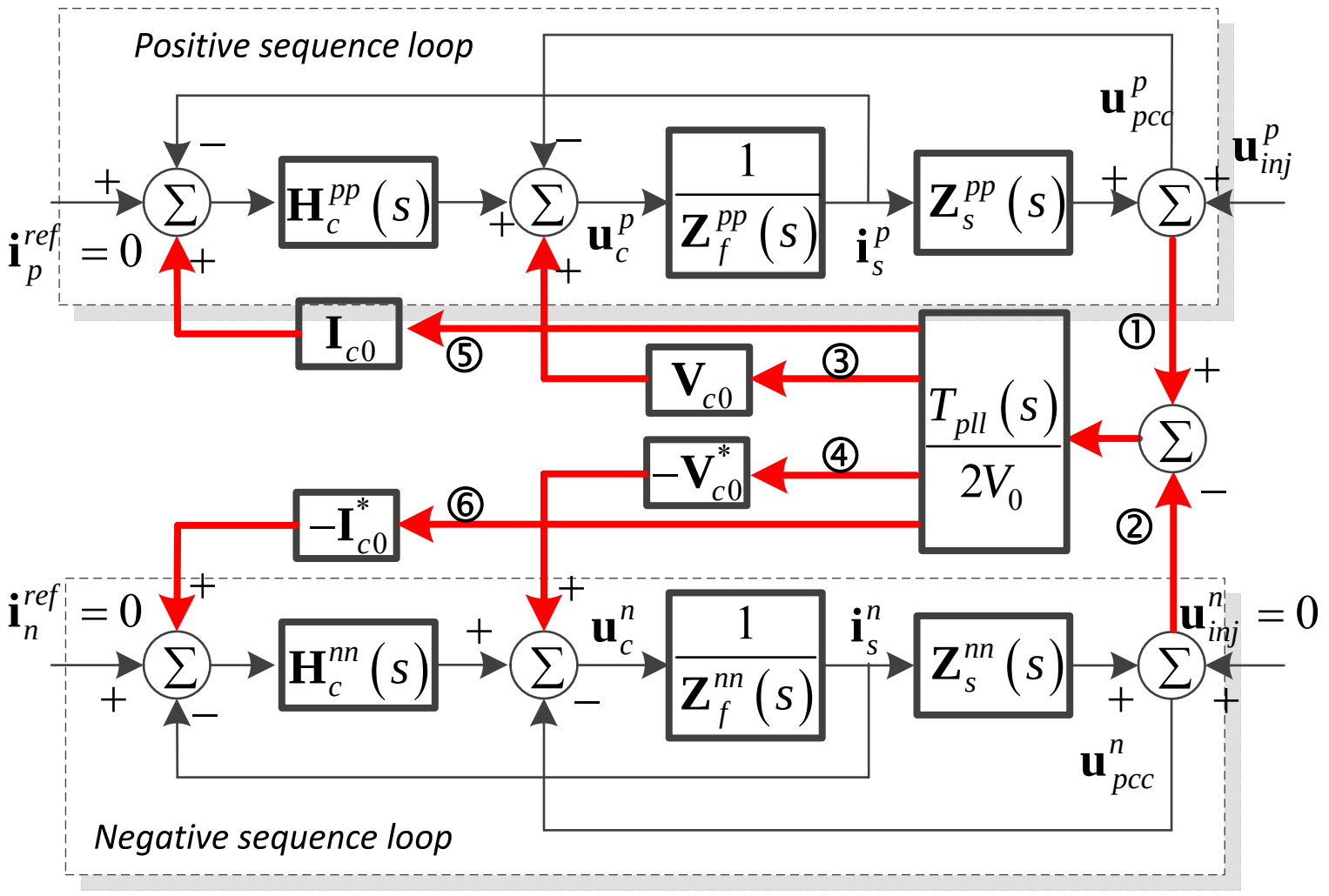

Fig. 3 System blocks diagram of grid-tied VSC in sequence domain

From Fig.3 we can clearly identify that the positive and negative sequence loops are coupled through six paths, all caused by the PLL loop gain $T_{pll}(s)$. Considering different paths will give sequence models with different accuracies, e.g.:

Case 1: Neglects all the paths, gives a simplest model with positive and negative sequence decoupled. Although this may not be effective for stability analysis directly, it is useful for identifying the intrinsic properties in the grid-VSC system, such as *intrinsic oscillation point* [22], and the coupling effects of PLL can be introduced as additional damping sources to the intrinsic point [22].

Case 2: Consider paths ①③⑤and②④⑥ separately. This leads to another decoupled sequence model as shown in [7]. The loop impedance from perturbation voltage $\mathbf{u}_{inj}^{p}$ to the current response $\mathbf{i}_{s}^{p}$ can be calculated directly from Fig.3, is $1/\mathbf{Y}_{L}^{pp}+\mathbf{Z}_{S}^{pp}$. Similarly, the negative counterpart is $1/\mathbf{Y}_L^{nn}+\mathbf{Z}_S^{nn}$. This condition is satisfied if the grid is relatively strong. (See the proof in (11) and (12)).

The foregoing analysis gives two decoupled sequence impedance models of grid-tied VSC, both of which are SISO systems. However, both of them neglect the coupling to some extent. The following section will derive an accurate SISO-model of the system where the positive and negative loops are viewed as subsystems of a closed-loop system.

B. Accurate and reduced SISO models of grid-tied VSC

In this section, we regard the VSC and grid as a closed-loop system instead of VSC and grid subsystems. Due to the linearity, positive and negative sequence perturbation to the closed-loop system can be analyzed separately. Taking the positive sequence perturbation as an example, the closed-loop system was drawn in Fig.4.

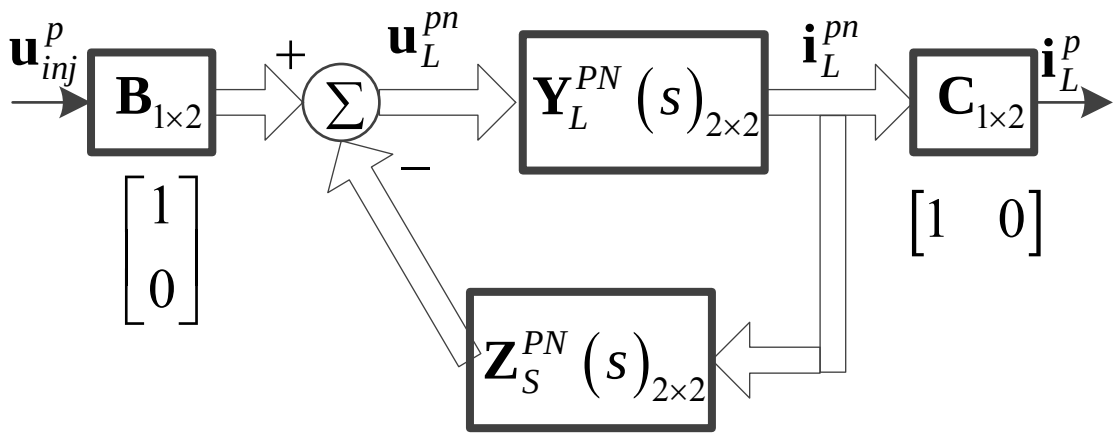

Fig. 4 The closed-loop representation of grid-tied VSC system

Solving the linear system in Fig.4 yields the positive sequence loop impedance:

$$\mathbf{Z}_{loop}^p(s)=-\frac{\mathbf{u}_{inj}^p}{\mathbf{i}_L^p}=\frac{1}{\mathbf{C}\left(\mathbf{Z}_L^{PN}(s)+\mathbf{Z}_S^{PN}(s)\right)^{-1}\mathbf{B}} \quad (14)$$

Where, $\mathbf{Z}_L^{PN}(s)=\left(\mathbf{Y}_L^{PN}(s)\right)^{-1}$.

Substitute (10) and (11) into (14) yields:

$$\mathbf{Z}_{loop}^p(s)=\mathbf{Z}_S^{pp}+\left(\mathbf{Z}_L^{pp}-\frac{\mathbf{Z}_L^{pn}\mathbf{Z}_L^{np}}{\mathbf{Z}_S^{nn}+\mathbf{Z}_L^{nn}}\right)=\mathbf{Z}_S^{pp}+\frac{1}{\mathbf{Y}_{L-eq}^{pp}} \quad (15)$$

Applying this method to find the negative sequence loop impedance by replacing the matrix $\mathbf{B}=[0\quad 1]^T$, $\mathbf{C}=[0\quad 1]$, and $\mathbf{u}_{inj}^p\rightarrow\mathbf{u}_{inj}^n$, yields:

$$\mathbf{Z}_{loop}^n(s)=\mathbf{Z}_S^{nn}+\left(\mathbf{Z}_L^{nn}-\frac{\mathbf{Z}_L^{pn}\mathbf{Z}_L^{np}}{\mathbf{Z}_S^{pp}+\mathbf{Z}_L^{pp}}\right)=\mathbf{Z}_S^{nn}+\frac{1}{\mathbf{Y}_{L-eq}^{nn}} \quad (16)$$

(15) and (16) are two SISO systems. More importantly, for each SISO system, the sequence couplings are accurately included. A physical interpretation is that the negative sequence circuit in Fig.2 is augmented into the positive sequence network (and vice versa) by solving the mutually dependent voltage-controlled-current-source, which introduced the couplings (Fig.3).

Further, considering a relatively strong grid, $\left|\mathbf{Z}_S^{pp}\right|\ll\left|\mathbf{Z}_L^{pp}\right|,\forall\omega$ and $\left|\mathbf{Z}_S^{nn}\right|\ll\left|\mathbf{Z}_L^{nn}\right|,\forall\omega$, (15) and (16) can be reduced to:

$$\tilde{\mathbf{Z}}^{p}_{loop}(s) = \mathbf{Z}^{pp}_{S} + \frac{\det\left|\mathbf{Z}^{PN}_{L}(s)\right|}{\mathbf{Z}^{nn}_{L}} = \mathbf{Z}^{pp}_{S} + \frac{1}{\mathbf{Y}^{pp}_{L}(s)} \quad (17)$$

$$\tilde{\mathbf{Z}}^{n}_{loop}(s) = \mathbf{Z}^{nn}_{S} + \frac{\det\left|\mathbf{Z}^{PN}_{L}(s)\right|}{\mathbf{Z}^{pp}_{L}} = \mathbf{Z}^{nn}_{S} + \frac{1}{\mathbf{Y}^{nn}_{L}(s)} \quad (18)$$

(17) and (18) is equivalent to the loop impedances (Section III.A Case 2) obtained directly from control blocks in Fig.3.

C. Comparison of analytical models with impedance measurements

There are two analytical models to be compared, i.e. the accurate model (model A) from (15) and (16), and the reduced model (model B) from (17) and (18). Impedance measurements are carried out in PSCAD/EMTDC with VSC and grid model shown in Fig.1shown (circuit parameters are in the Appendix A). The multi-run module in PSCAD is used, during each run, a single tone harmonic voltage is injected into the grid. The frequency injected is from 0 Hz to 100 Hz with an increment of 2Hz. The sampling frequency and sampling window used for Fourier analysis are 1 kHz and 0.5 sec respectively. All the figures and data are post-processed in Matlab.

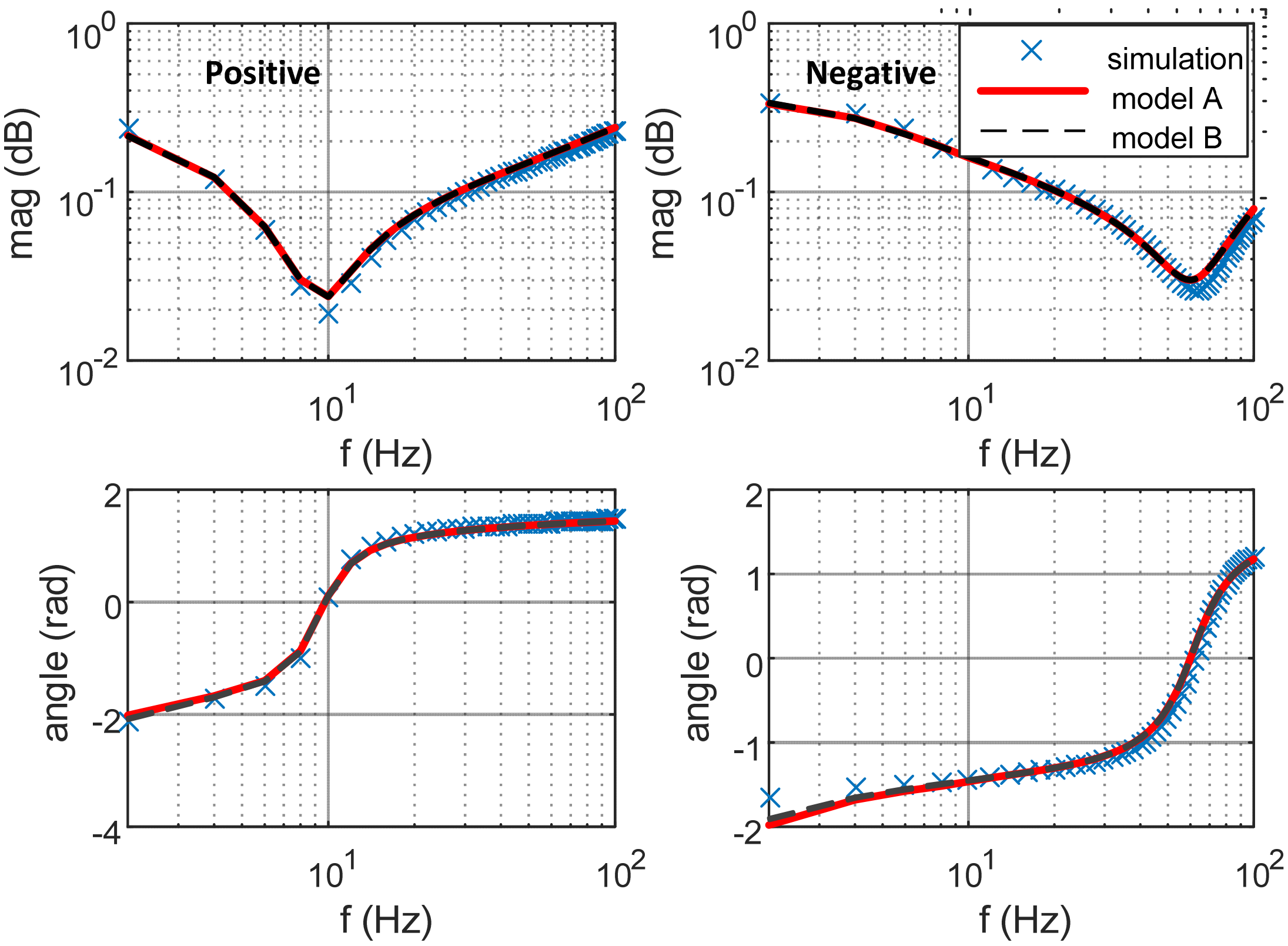

(a) Closed-loop impedance plots (condition 1)
(SCR=4, CC=200Hz, PLL=5Hz, nominal current (flow out), horizontal axis is frequency in *dq* frame, vertical axis for upper plots are magnitude and lower plots are phase angle)

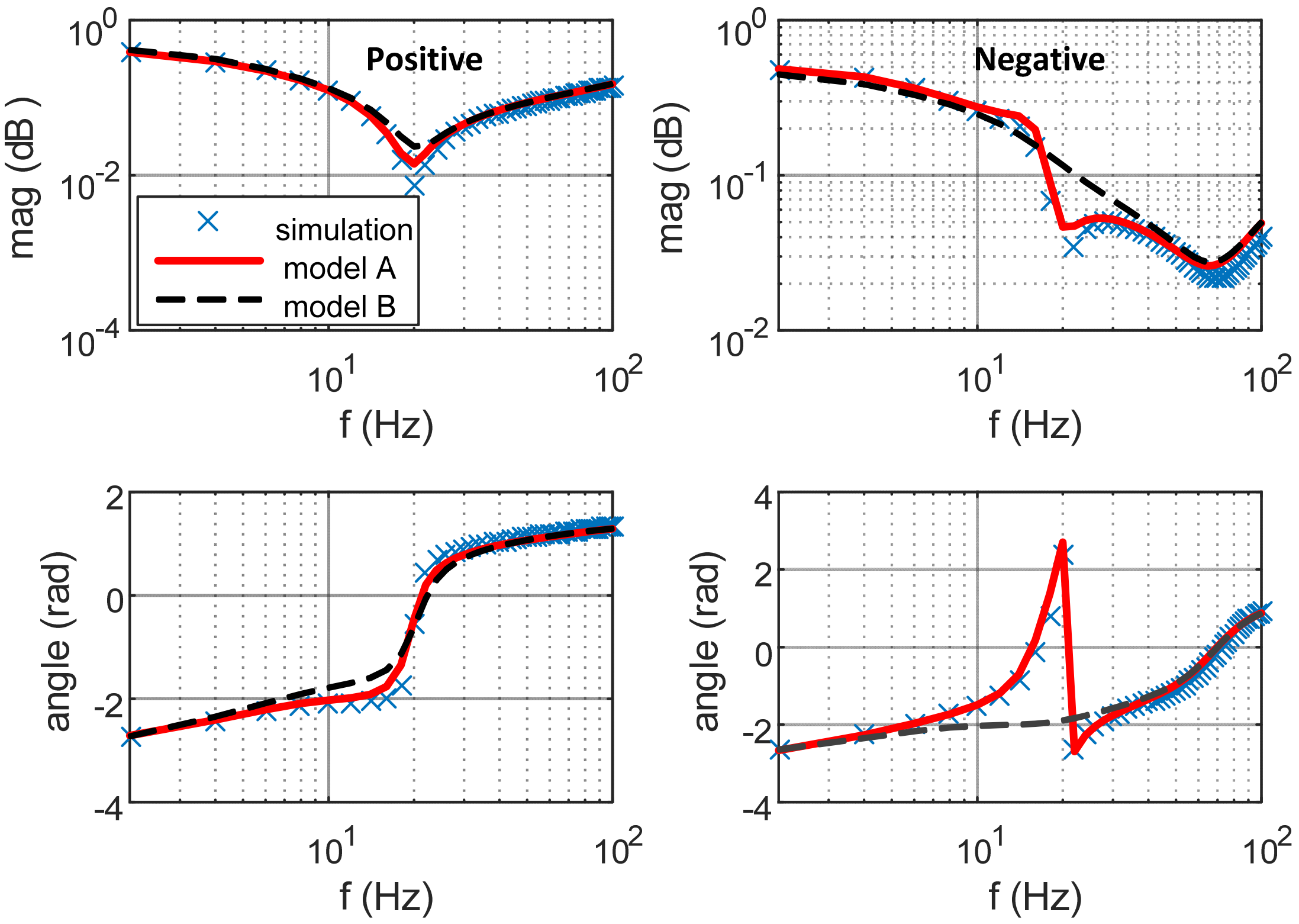

(b) Closed-loop impedance plots (condition 2)

(SCR=8, CC=200Hz, PLL=100Hz, nominal current (flow out), horizontal axis is frequency in *dq* frame, vertical axis for upper plots are magnitude and lower plots are phase angle)

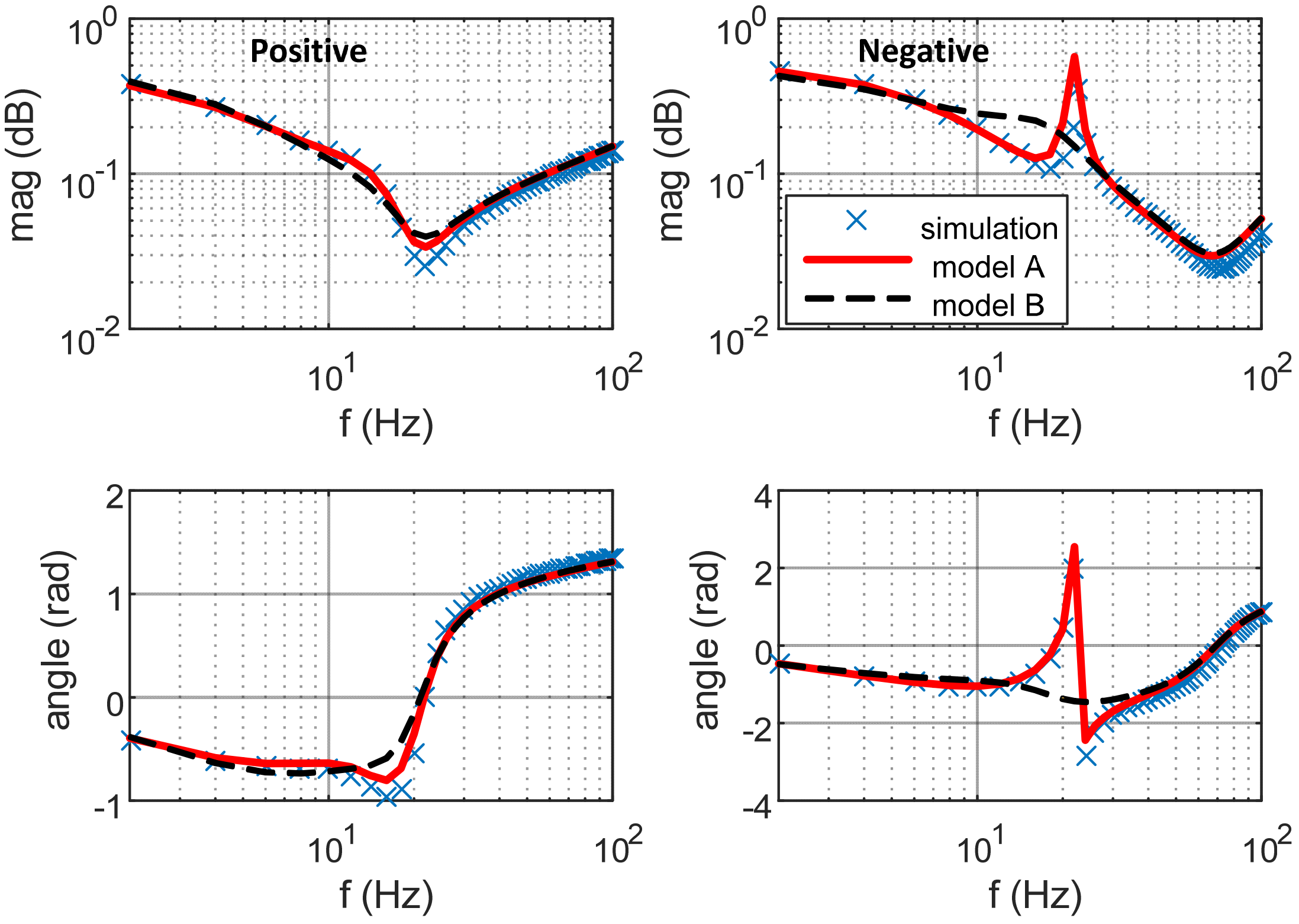

(c) Closed-loop impedance plots (condition 3)

(SCR=8, CC=200Hz, PLL=100Hz, nominal current (flow in), horizontal axis is frequency in *dq* frame, vertical axis for upper plots are magnitude and lower plots are phase angle)

Fig. 5 Comparison of analytical SISO impedance models with simulation measurements

Fig.5 (a) illustrates that both accurate and reduced models achieve a good match with the measured impedances under a small PLL bandwidth condition. However, if PLL bandwidth was increased to a relatively large value, the shapes of reduced model differ from the measurements, particularly for the negative sequence impedances as shown in Fig.5 (b). On the contrary, the accurate SISO model can track the measured impedances even at the transition sections in Fig.5 (b). Fig.5 (c) plots the impedances with inversed power flow direction. It also proves that the accurate model is superior to the reduced model in capturing the transition of impedance characteristics. Note that the bandwidths for PLL and current controller under a specific grid condition are carefully chosen in accordance to the analysis in [22] to assure a stable operational point.

From the above comparisons, we can identify that the PLL has a great impact on the impedance within its control bandwidth (normally at low frequencies). The accurate model is capable of capturing this characteristic change but the reduced model cannot. However, the reduced model can still be effective if these model discrepancies have negligible effects on small signal stability. Therefore, the effects of model discrepancies on the small signal stability should be further justified. This is performed in the next section.

## IV. Small signal stability analysis

This section intends to analyze the validity of proposed SISO models in predicting small signal stability. Benefit from the SISO properties, the proposed model can be used in combination with classic Nyquist criterion (NC) [23] for stability analysis.

### A. Numerical stability analysis

Three combinations of numerical plots are considered in this part:

1) Model A SISO with (NC). (To be compared).
2) Model B SISO with (NC). (To be compared).
3) Model C MIMO ((10) and (11)) with (GNC). (Reference model).

In 1) the locus of minor loop gains are $\mathbf{l}_p(s)=\mathbf{Z}_S^{pp}\mathbf{Y}_{L-eq}^{pp}(s)$ and $\mathbf{l}_n(s)=\mathbf{Z}_S^{nn}\mathbf{Y}_{L-eq}^{nn}(s)$ in accordance with (15) and (16). In 2) the minor loop gains are $\tilde{\mathbf{l}}_p(s)=\mathbf{Z}_S^{pp}\mathbf{Y}_L^{pp}(s)$ and $\tilde{\mathbf{l}}_n(s)=\mathbf{Z}_S^{nn}\mathbf{Y}_L^{nn}(s)$ in accordance with (17) and (18). In 3) the eigenvalue loci can be calculated from $\det\left|\mathbf{L}(s)-\mathbf{Z}_S^{PN}\mathbf{Y}_L^{PN}(s)\right|$, where $\mathbf{L}(s)=diag_{2\times2}(l_1,l_2)$. Note that the models above are generally complex transfer functions, thus the locus for the negative frequencies are not the conjugated of the locus of positive frequencies [16]. However, the complex transfer functions for SISO systems have the property $\mathbf{l}_p(-j\omega)=\left(\mathbf{l}_n(j\omega)\right)^*$, thus the negative frequency plots can be obtained by conjugating the negative sequence locus.

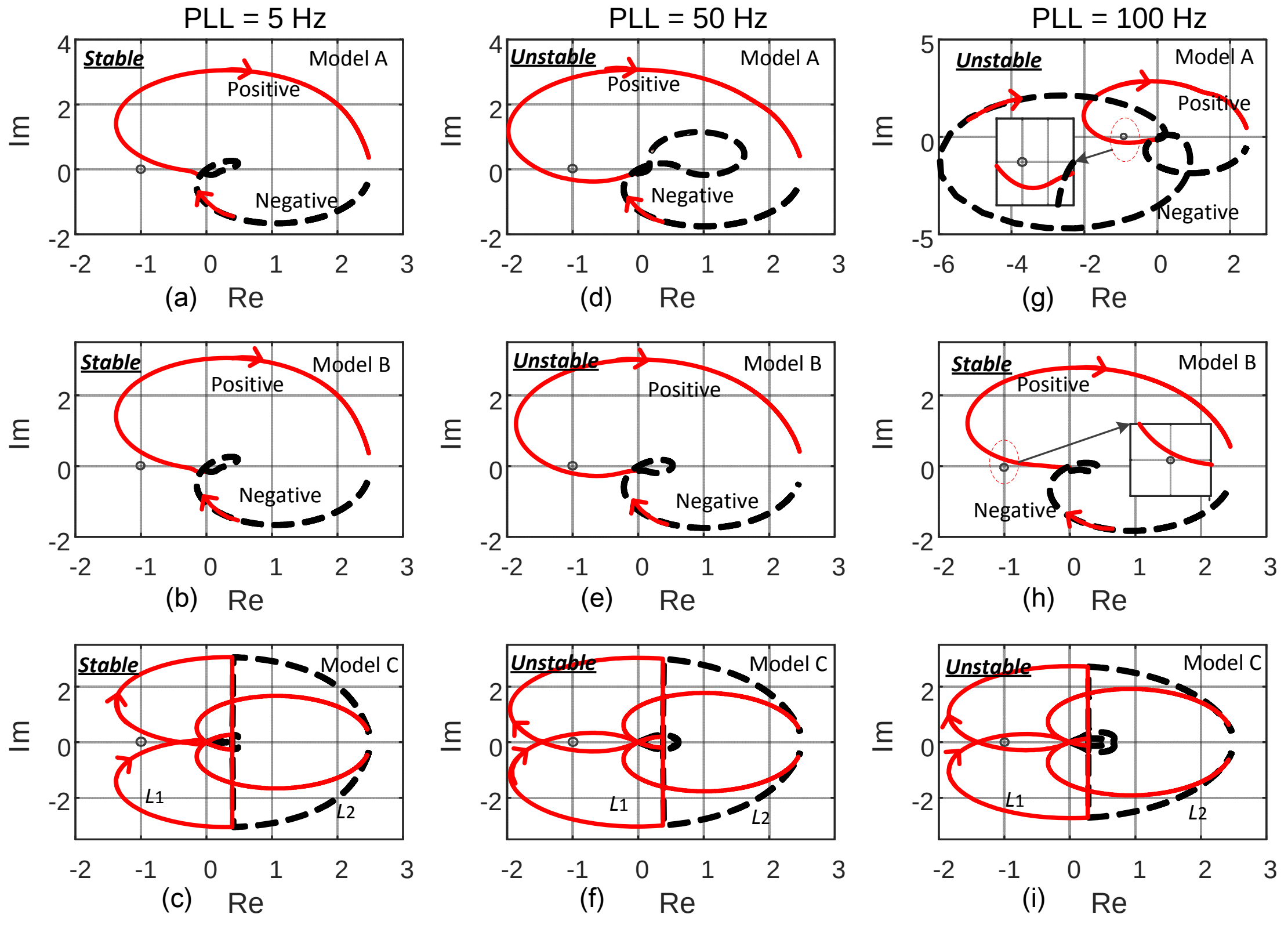

Fig. 6 Numerical stability comparisons

(Grid SCR is 4. Current controller bandwidth is 200 Hz. The VSC is loaded with nominal current (flow out). The PLL bandwidth is chosen as 5 Hz, 50 Hz and 100 Hz respectively)

As shown in Fig.6, the Model A matches the stability results of Model C in all PLL bandwidth conditions, i.e. from the first column to the third column of fig.6 the stability results are stable, unstable and unstable, which proves the validity of Model A. On the other hand, the Model B predicts the wrong stability results under large PLL bandwidth conditions, as shown in Fig.6 (h) compared with (g) and (i).

It is intuitively illustrated in Fig.7 (g) and (h) that the negative sequence locus of Model B differs largely from Model A This was also identified in the foregoing analysis in Fig.5. Moreover, although the changes in the positive sequence locus of Model B were small compared with Model A, its consequences on stability are large. In Fig.7 (h), the positive sequence locus does not encircle point at (-1,0j), whereas the positive sequence locus in Fig.7 (g) does, therefore Model B fails to predict the stability accurately. This finding can also be justified by the passivity of positive and negative loop impedances of Model A and B.

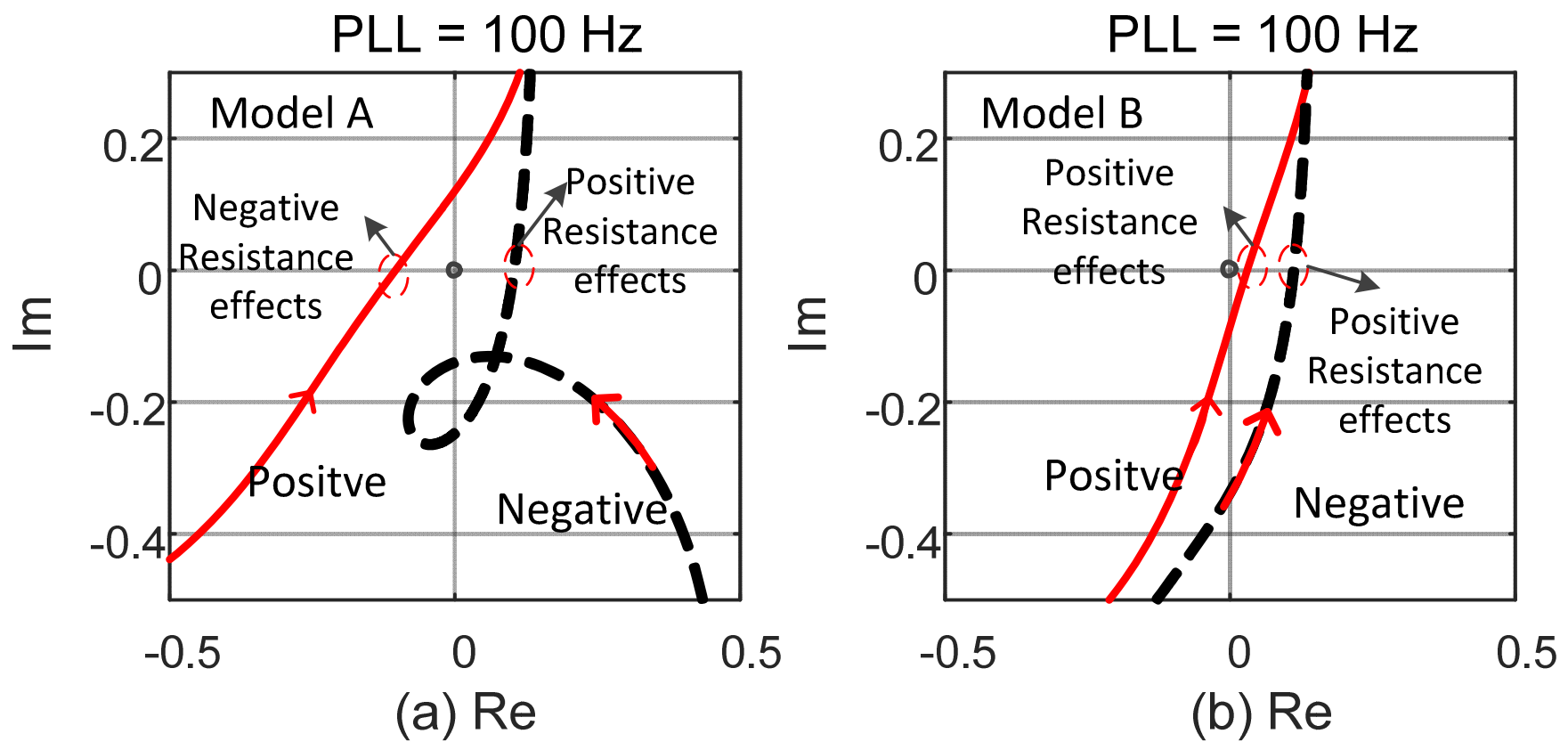

Fig. 7 passivity analysis

As shown in Fig.7 (a), the positive sequence component of Model A has negative resistance effects as the real axis is being crossed, whereas the negative component of Model A has positive resistance effects. On the other hand, in Fig.7 (b), both the positive and negative components of Model B exhibit positive resistance effects. A positive resistance effect implies a stable resonance, therefore the passivity analysis [24] gives the same stability results as the case in Fig.6 (g) and (h).

B. Simulation study

By varying PLL bandwidth, the marginally stable condition of grid-tied VSC was found based on Model A, as shown in Fig.8, which is 18 Hz. And the positive and negative sequence oscillation frequencies can be determined directly from the plots, which is approximately 10 Hz and 60 Hz respectively (referred to *dq* frame). However, due to positive resistance effects at negative sequence resonance (Fig.7 a), it is supposed to be stable. Therefore, only the positive sequence oscillation at 10 Hz is expected. The parameter values corresponding to marginally stable condition are then applied in the simulation model (Fig.1) built in PSCAD/EMTDC to perform time domain verification.

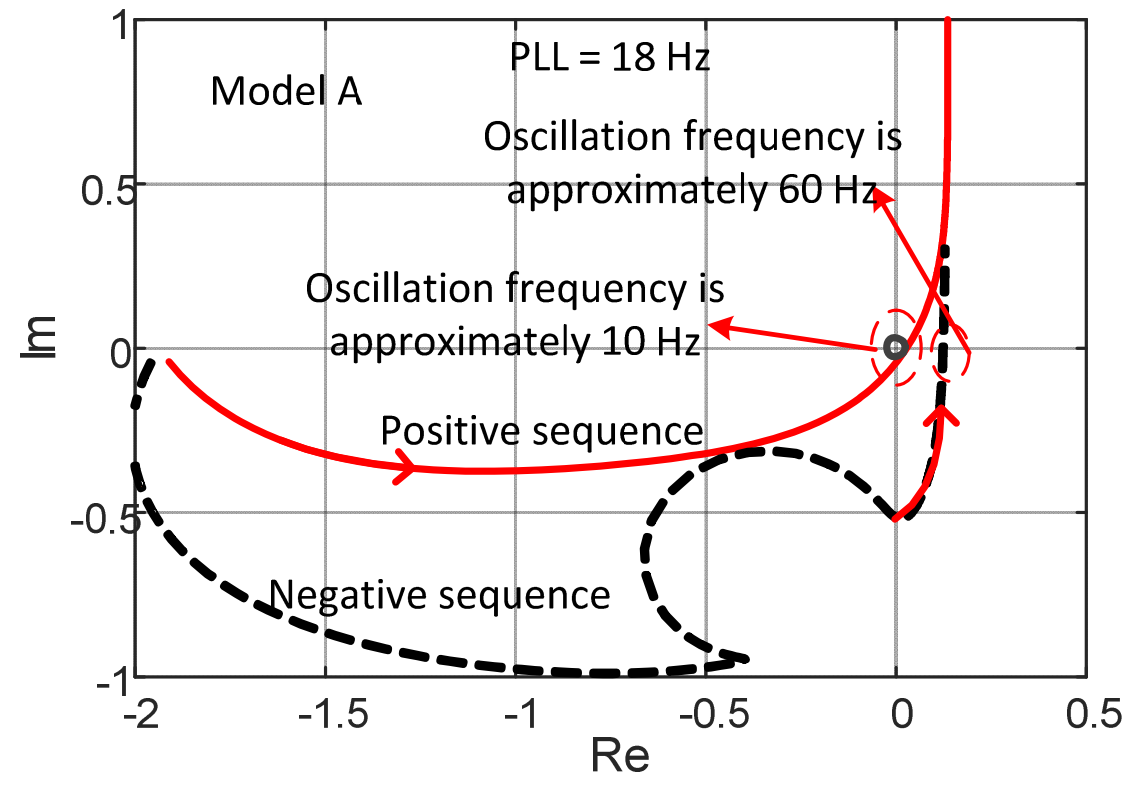

(a) Loop impedances plots

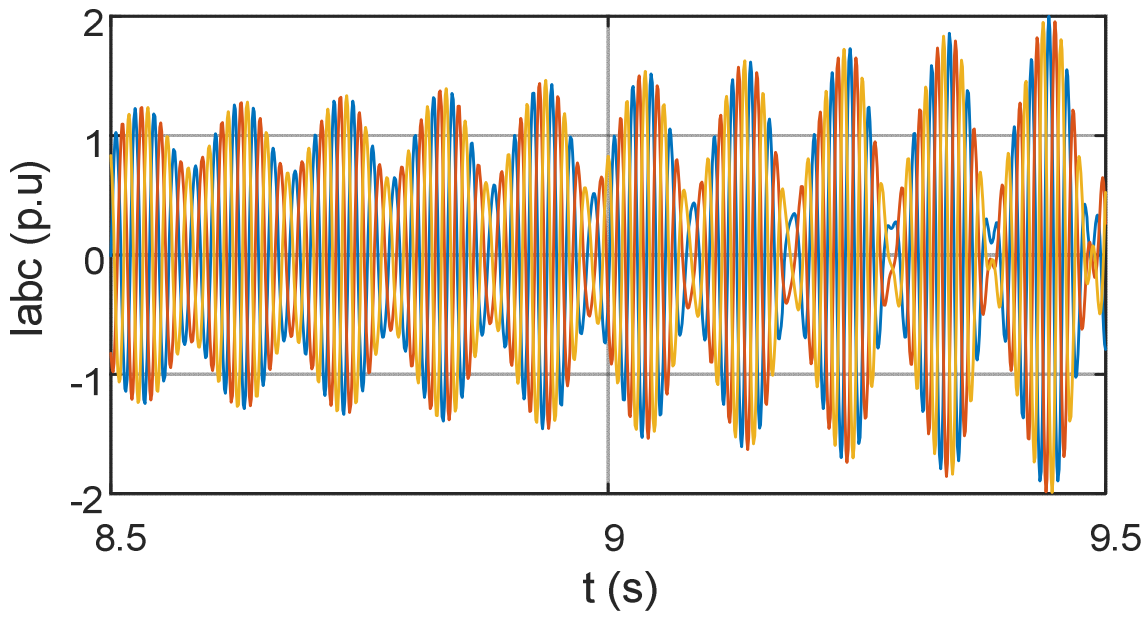

(b) Time domain simulation

(Before 2 s, the PLL bandwidth is set to 5 Hz to achieve a stable operational point, afterwards the PLL bandwidth is set to 20 Hz. The oscillation is observable after several seconds)

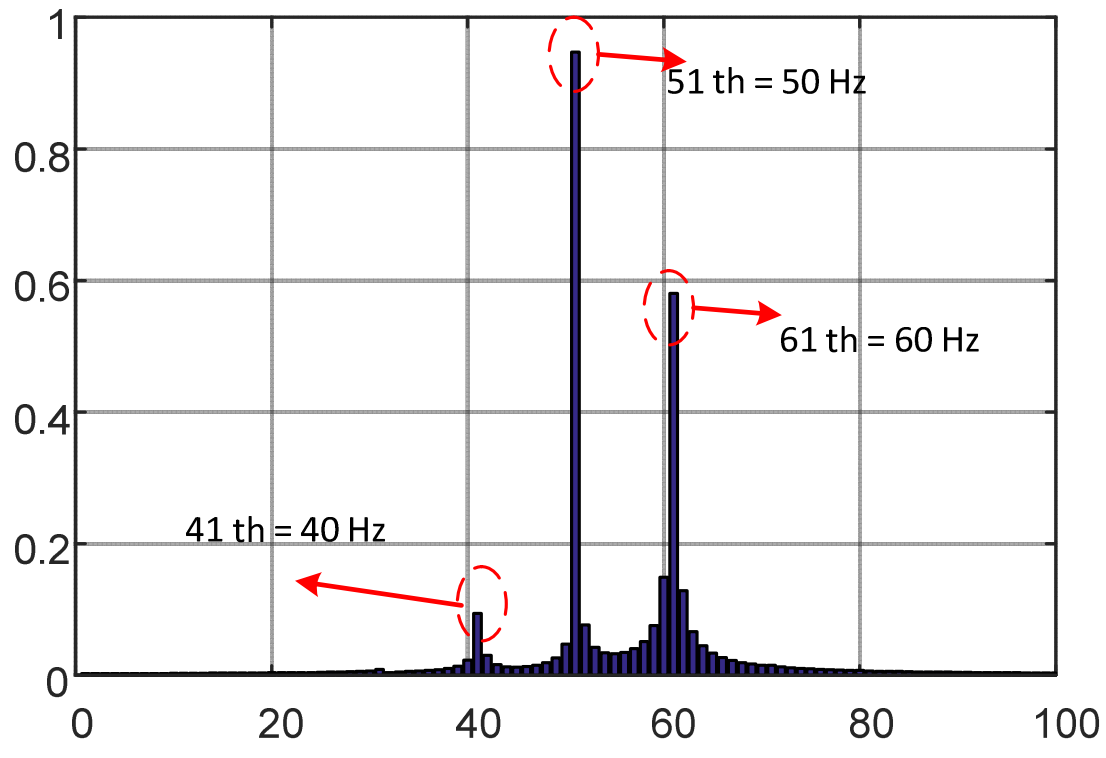

(c) Fourier analysis of phase current

(sampling rate is 1 kHz, sampling window is 1 sec)

Fig. 8 marginal stable simulations

(Current controller bandwidth is 200 Hz, SCR is 4, VSC is loaded with nominal current (flow out))

Fig.8 (b) presents that the system is small signal unstable as predicted by Model A in Fig.8 (a). In Fig.8 (c), the mirror frequencies 40 Hz and 60 Hz are originated from the positive sequence oscillation in *dq* frame, which is 10 Hz specified by Fig.8 (a). The oscillatory behavior shown in Fig.8 (b) is similar to the field measurements of grid-tied photovoltaic inverter systems in [25].

## V. CONCLUSIONS

This paper investigated the SISO sequence impedance models of grid-tied VSC system. In the perspective of physical understanding, the sequence coupling paths were revealed intuitively from the established sequence block diagrams. Subsequently, two SISO models have been proposed, one was derived from the strong grid assumption (refer to reduced SISO model), i.e. the positive and negative system loops were considered separately. The other one was derived from the closed-loop analysis of grid-tied VSC system (refer to accurate SISO model), i.e. the negative sequence loop was augmented into the positive sequence loop if positive sequence perturbation was analyzed. By means of numerical and time domain analysis, the accurate model gives identical stability analysis results as the matrix impedance models, whereas the reduced model can lead to wrong stability results if PLL bandwidth was relatively large. A consequential benefit from the simplicity of the proposed SISO model is that, the stability can be assessed by the classic Nyquist criterion or passivity analysis. In addition, the impedance measurements can be simple compared with *dq* impedance or matrix sequence impedance.

In theory, the proposed SISO model is effective for any single frequency coupling system caused by *dq* asymmetry, e.g. VSC includes power controller or dc voltage controller etc., however, this should be justified in the future works.

ACKNOWLEDGEMENT

This paper and its related research are supported by grants from the Power Electronics Science and Education Development Program of Delta Environmental & Educational Foundation (DREM2016005).

APPENDIX:

A. *Circuit parameters used in stability analysis and simulations*

TABLE.I Circuit parameters of the grid-tied VSC system

| Name | Values | Name | Values |
|---|---|---|---|
| Nominal power | 2 MVA | filter impedance | (0.025+0.1j) *p.u.* |
| Nominal voltage | 0.69kV | Current reference | (0.5 +0j) *p.u.* |
| DC link voltage | 1.1kV | Grid impedance | reciprocal of SCR |
| Switching frequency | 2.4 kHz | Grid X/R ratio | 5 |
| Fundamental frequency | 50Hz | | |